\documentclass[fleqn,twoside]{article}
\usepackage{amsmath,espcrc2,graphicx}
\title{Matching QCD and HQET at three loops}
\author{A.G.~Grozin%
\address{Institut f\"ur Theoretische Teilchenphysik,
Karlsruher Institut f\"ur Technologie}}
       
\begin{document}
\begin{abstract}
QCD/HQET matching for the heavy-quark field~\cite{G:10}
and heavy--light quark currents~\cite{BGMPSS:10}
with three-loop accuracy is discussed.
\vspace{1pc}
\end{abstract}

\maketitle

\section{Heavy-quark field}
\label{S:Field}

QCD problems with a single heavy quark $Q$
can be treated in a simpler effective theory --- HQET,
if there exists a 4-velocity $v$ such that
the heavy-quark momentum is $p=mv+k$
($m$ is the on-shell mass)
and the characteristic residual momentum is small: $k\ll m$.
QCD operators can be written as series in $1/m$ via HQET operators;
the coefficients in these series are determined by matching
on-shell matrix elements in both theories.

At the tree level, the heavy-quark field $Q$
is related to the corresponding HQET field $Q_v$
(satisfying $\rlap/v Q_v=Q_v$) by~\cite{Lee:91,KT:91}
\begin{align}
&Q(x) = e^{-i m v\cdot x}
\left(1 + \frac{i \rlap{\hspace{0.2em}/}D_\bot}{2m} + \cdots\right)
Q_v(x)\,,
\nonumber\\
&D^\mu_\bot=D^\mu-v^\mu v\cdot D\,.
\label{tree}
\end{align}
The matrix elements of the bare fields
between the on-shell quark with momentum $p=mv+k$ and the vacuum
in both theories are given
by the on-shell wave-function renormalization constants:
\begin{align}
&{<}0|Q_0|Q(p){>} = \left(Z_Q^{\text{os}}\right)^{1/2} u(p)\,,
\nonumber\\
&{<}0|Q_{v0}|Q(p){>} = \left(\tilde{Z}_Q^{\text{os}}\right)^{1/2} u_v(k)
\label{onshell}
\end{align}
(HQET renormalization constants are denoted by $\tilde{Z}$).
The Dirac spinors are related by the Foldy--Wouthuysen transformation
\begin{equation*}
u(mv+k) = \left[1 + \frac{\rlap/k}{2m}
+ \mathcal{O}\left(\frac{k^2}{m^2}\right) \right] u_v(k)\,.
\end{equation*}
Therefore, the bare fields are related by
\begin{align}
Q_0(x) = e^{-i m v\cdot x} \biggl[& z_0^{1/2}
\left(1 + \frac{i \rlap{\hspace{0.2em}/}D_\bot}{2m}\right) Q_{v0}(x)
\nonumber\\
&{} + \mathcal{O}\left(\frac{1}{m^2}\right) \biggr]\,,
\label{bare}
\end{align}
where the bare matching coefficient is
\begin{equation}
z_0 = \frac{Z_Q^{\text{os}}(g_0^{(n_l+1)},a_0^{(n_l+1)})}%
{\tilde{Z}_Q^{\text{os}}(g_0^{(n_l)},a_0^{(n_l)})}
\label{z0}
\end{equation}
(we use the covariant gauge: the gauge-fixing term in the Lagrangian
is $-(\partial_\mu A_0^{a\mu})/(2a_0)$, and the free gluon propagator
is $(-i/p^2) (g_{\mu\nu} - (1-a_0) p_\mu p_\nu/p^2)$;
the number of flavours in QCD is $n_f=n_l+1$).
The $\mathcal{O}(1/m)$ matching coefficient in~(\ref{bare})
is equal to the leading one, $z_0$;
this reflexes the reparametrization invariance~\cite{LM:92}.
The $\overline{\text{MS}}$ renormalized fields are related
by the formula similar to~(\ref{bare}),
with the renormalized decoupling coefficient
\begin{equation}
z(\mu) =
\frac{\tilde{Z}_Q(\alpha_s^{(n_l)}(\mu),a^{(n_l)}(\mu))}%
{Z_Q(\alpha_s^{(n_l+1)}(\mu),a^{(n_l+1)}(\mu))} z_0\,.
\label{renorm}
\end{equation}

If there are no massive flavours except $Q$,
then $\tilde{Z}_Q^{\text{os}}=1$
because all loop corrections are scale-free.
The QCD on-shell renormalization constant $Z_Q^{\text{os}}$
contains the single scale $m$ in this case;
it has been calculated~\cite{MR:00} up to three loops.
The three-loop $\overline{\text{MS}}$ anomalous dimensions
of $Q_v$~\cite{MR:00,CG:03} and $Q$~\cite{T:82,LV:93} are also known.
We have to express all three quantities
$Z_Q^{\text{os}}(g_0^{(n_l+1)},a_0^{(n_l+1)})$,
$Z_Q(\alpha_s^{(n_l+1)}(\mu),a^{(n_l+1)}(\mu))$,
$\tilde{Z}_Q(\alpha_s^{(n_l)}(\mu),a^{(n_l)}(\mu))$
via the same variables, say,
$\alpha_s^{(n_l)}(\mu)$, $a^{(n_l)}(\mu)$,
see~\cite{CKS:98}.
The explicit result for the renormalized matching coefficient $z(\mu)$
can be found in~\cite{G:10}.
Gauge dependence first appears at three loops,
as in $Z_Q^{\text{os}}$~\cite{MR:00}.
The requirement of finiteness
of the renormalized matching coefficient~(\ref{renorm})
at $\varepsilon\to0$ has allowed the authors of~\cite{MR:00}
to extract $\tilde{Z}_Q$ from their result for $Z_Q^{\text{os}}$.

In the large-$\beta_0$ limit
(see Chapter~8 in~\cite{G:04} for a pedagogical introduction):
\begin{align}
z(\mu) ={}& 1 + \int_0^\beta \frac{d\beta}{\beta}
\left(\frac{\gamma(\beta)}{2\beta} - \frac{\gamma_0}{2\beta_0}\right)
\nonumber\\
&{} + \frac{1}{\beta_0} \int_0^\infty du\,e^{-u/\beta} S(u)
+ \mathcal{O}\left(\frac{1}{\beta_0^2}\right)\,,
\label{largeb0}
\end{align}
where $\beta=\beta_0\alpha_s/(4\pi)$,
$\gamma=\gamma_0\alpha_s/(4\pi)+\cdots$
(differences of $n_l$-flavour and $(n_l+1)$-flavour quantities
can be neglected at the $1/\beta_0$ order).
The difference of the QCD and HQET anomalous dimensions
$\gamma=\gamma_Q-\tilde{\gamma}_Q$
(it is gauge invariant at this order)
and the Borel image $S(u)$ are~\cite{BG:95,NS:95,G:04}
\begin{align}
&\gamma(\beta) = - 2 \frac{\beta}{\beta_0} F(-\beta,0) = {}
\nonumber\\
&2 C_F \frac{\beta}{\beta_0}
\frac{(1+\beta) (1+\frac{2}{3}\beta)}%
{B(2+\beta,2+\beta) \Gamma(3+\beta) \Gamma(1-\beta)}\,,
\nonumber\\
&S(u) = \frac{F(0,u)-F(0,0)}{u} = {}
\label{struct}\\
& - 6 C_F \biggl[ e^{(L+5/3)u}
\frac{\Gamma(u) \Gamma(1-2u)}{\Gamma(3-u)} (1-u^2)
- \frac{1}{2u} \biggr]\,.
\nonumber
\end{align}
This Borel image has infrared renormalon poles at each positive
half-integer $u$ and at $u=2$.
Therefore, the integral in~(\ref{largeb0}) is not well defined.
Comparing its residue at the leading pole $u=1/2$
with the residue of the static-quark self-energy at its
ultraviolet pole $u=1/2$~\cite{BB:94},
we can express the renormalon ambiguity of $z(\mu)$ as
\begin{equation}
\Delta z(\mu) = \frac{3}{2} \frac{\Delta\bar{\Lambda}}{m}
\label{ambig}
\end{equation}
($\bar{\Lambda}$ is the ground-state meson residual energy).
The matching coefficient is gauge invariant at the order $1/\beta_0$.
Expanding $\gamma(\beta)$ and $S(u)$ and integrating,
we obtain confirm the contributions with the highest power of $n_l$
in each term in our three-loop result,
and predict such a contribution at $\alpha_s^4$.

Numerically, in the Landau gauge at $n_l=4$
\begin{align}
&z(m) = 1 - \frac{4}{3} \frac{\alpha_s^{(4)}(m)}{\pi}
\nonumber\\
&{} - (16.6629 - 4.5421)
\left(\frac{\alpha_s^{(4)}(m)}{\pi}\right)^2
\nonumber\\
&{} - (153.4076 + 42.6271 - 61.5397)
\left(\frac{\alpha_s^{(4)}(m)}{\pi}\right)^3
\nonumber\\
&{} - (1953.4013 + \cdots)
\left(\frac{\alpha_s^{(4)}(m)}{\pi}\right)^4
+ \cdots
\nonumber\\
&{} = 1 - \frac{4}{3} \frac{\alpha_s^{(4)}(m)}{\pi}
- 12.1208
\left(\frac{\alpha_s^{(4)}(m)}{\pi}\right)^2
\nonumber\\
&{} - 134.4950
\left(\frac{\alpha_s^{(4)}(m)}{\pi}\right)^3
\nonumber\\
&{} - (1953.4013 + \cdots)
\left(\frac{\alpha_s^{(4)}(m)}{\pi}\right)^4
+ \cdots
\label{numeric}
\end{align}
($\beta_0$ is for $n_l=4$ flavours).
Naive nonabelianization~\cite{BG:95} works rather well at two
and three loops.
Numerical convergence of the series is very poor;
this is related to the infrared renormalon at $u=1/2$.

Now let us consider the relation
between the $\overline{\text{MS}}$ renormalized electron field in QED
and the Bloch--Nordsieck electron field.
The bare matching coefficient $z_0=Z_\psi^{\text{os}}$
is gauge invariant to all orders, see~\cite{MR:00}.
In the Bloch-Nordsieck model, due to exponentiation,
$\log\tilde{Z}_\psi=(3-a^{(0)})\alpha^{(0)}/(4\pi\varepsilon)$
(where the 0-flavour $\alpha^{(0)}$
is equal to the on-shell $\alpha\approx1/137$).
In the full QED,
$\log Z_\psi=a^{(1)}\alpha^{(1)}/(4\pi\varepsilon)
+(\text{gauge-invariant higher terms})$
(see~\cite{G:10} for the proof;
this has been demonstrated up to four loops
by the direct calculation~\cite{CR:00}).
The gauge dependence cancels in $\log(\tilde{Z}_\psi/Z_\psi)$
because of the QED decoupling relation
$a^{(1)}\alpha^{(1)}=a^{(0)}\alpha^{(0)}$.
Therefore, the renormalized matching coefficient $z(\mu)$ in QED
is gauge invariant to all orders.
The three-loop result is presented in~\cite{G:10}.

\section{Heavy--light currents}
\label{S:Currents}

Now we shall consider~\cite{BGMPSS:10} $\overline{\text{MS}}$ renormalized
heavy--light QCD quark currents
\begin{equation}
j(\mu) = Z_j^{-1}(\mu) j_0\,,\quad
j_0 = \bar{q}_0 \Gamma Q_0\,,
\label{Intro:j}
\end{equation}
where $\Gamma$ is a Dirac matrix.
They can be expressed via operators in HQET
\begin{align}
j(\mu) ={}& C_\Gamma(\mu) \tilde{\jmath}(\mu)
+ \frac{1}{2m} \sum_i B_i(\mu) O_i(\mu)
\nonumber\\
&{} + \mathcal{O}\left(\frac{1}{m^2}\right)\,,
\label{Intro:match}
\end{align}
where
\begin{equation}
\tilde{\jmath}(\mu) = \tilde{Z}_j^{-1}(\mu) \tilde{\jmath}_0\,,\quad
\tilde{\jmath}_0 = \bar{q}_0 \Gamma Q_{v0}\,,
\label{Intro:jtilde}
\end{equation}
and $O_i$ are dimension-4 HQET operators
with appropriate quantum numbers.
The leading-order matching coefficients $C_\Gamma$
have been calculated up to two loops~\cite{BG:95,G:98}.

There are 8 Dirac structures giving non-vanishing quark currents
in 4 dimensions:
\begin{align}
&\Gamma = 1\,,\quad
\rlap/v\,,\quad
\gamma_\bot^\alpha\,,\quad
\gamma_\bot^\alpha \rlap/v\,,
\label{Match:Gamma}\\
&\gamma_\bot^{[\alpha} \gamma_\bot^{\beta]}\,,\quad
\gamma_\bot^{[\alpha} \gamma_\bot^{\beta]} \rlap/v\,,\quad
\gamma_\bot^{[\alpha} \gamma_\bot^{\beta} \gamma_\bot^{\gamma]}\,,\quad
\gamma_\bot^{[\alpha} \gamma_\bot^{\beta} \gamma_\bot^{\gamma]} \rlap/v\,,
\nonumber
\end{align}
where $\gamma_\bot^\alpha = \gamma^\alpha - \rlap/v v^\alpha$.
The last four of them can be obtained from the first four
by multiplying by the 't~Hooft--Veltman $\gamma_5^{\text{HV}}$.
We are concerned with flavour non-singlet currents only,
therefore, we may also use the anticommuting $\gamma_5^{\text{AC}}$
(there is no anomaly).
The currents renormalized at a scale $\mu$
with different prescriptions for $\gamma_5$ 
are related by~\cite{LV:91}
\begin{align}
&\left(\bar{q} \gamma_5^{\text{AC}} Q\right)_\mu =
Z_P(\mu) \left(\bar{q} \gamma_5^{\text{HV}} Q\right)_\mu\,,
\label{Match:Larin}\\
&\left(\bar{q} \gamma_5^{\text{AC}} \gamma^\alpha Q\right)_\mu =
Z_A(\mu) \left(\bar{q} \gamma_5^{\text{HV}} \gamma^\alpha Q\right)_\mu\,,
\nonumber\\
&\bigl(\bar{q} \gamma_5^{\text{AC}} \gamma^{[\alpha} \gamma^{\beta]} Q\bigr)_\mu =
Z_T(\mu) \bigl(\bar{q} \gamma_5^{\text{HV}} \gamma^{[\alpha} \gamma^{\beta]} Q\bigr)_\mu\,,
\nonumber
\end{align}
where the finite renormalization constants $Z_{P,A,T}$
can be reconstructed from the differences of the anomalous dimensions
of the currents.
Multiplying $\Gamma$ by $\gamma_5^{\text{AC}}$
does not change the anomalous dimension.
In the case of $\Gamma=\gamma^{[\alpha} \gamma^{\beta]}$,
multiplying it by $\gamma_5^{\text{HV}}$
just permutes its components,
and also does not change the anomalous dimension,
therefore,
\begin{equation}
Z_T(\mu) = 1\,;
\label{Match:ZT}
\end{equation}
$Z_{P,A}(\mu)$ are known up to three loops~\cite{LV:91}.

The anomalous dimension of the HQET current~(\ref{Intro:jtilde})
does not depend on the Dirac structure $\Gamma$.
Therefore, there are no factors similar to $Z_{P,A}$ in HQET.
Multiplying $\Gamma$ by $\gamma_5^{\text{AC}}$
does not change the matching coefficient.
Therefore, the matching coefficients for the currents
in the second row of~(\ref{Match:Gamma})
can be obtained from those for the first row.
In the $v$ rest frame
\begin{align}
Z_P(\mu) &{}=
\frac{C_{\gamma_5^{\text{AC}}}(\mu)}%
{C_{\gamma_5^{\text{HV}}}(\mu)} =
\frac{C_1(\mu)}%
{C_{\gamma^0 \gamma^1 \gamma^2 \gamma^3}(\mu)}\,,
\nonumber\\
Z_A(\mu) &{}=
\frac{C_{\gamma_5^{\text{AC}} \gamma^0}(\mu)}%
{C_{\gamma_5^{\text{HV}} \gamma^0}(\mu)} =
\frac{C_{\gamma^0}(\mu)}%
{C_{\gamma^1 \gamma^2 \gamma^3}(\mu)}
\nonumber\\
&{}=
\frac{C_{\gamma_5^{\text{AC}} \gamma^3}(\mu)}%
{C_{\gamma_5^{\text{HV}} \gamma^3}(\mu)} =
\frac{C_{\gamma^3}(\mu)}%
{C_{\gamma^0 \gamma^1 \gamma^2}(\mu)}\,,
\nonumber\\
Z_T(\mu) &{}=
\frac{C_{\gamma_5^{\text{AC}} \gamma^0 \gamma^1}(\mu)}%
{C_{\gamma_5^{\text{HV}} \gamma^0 \gamma^1}(\mu)} =
\frac{C_{\gamma^0 \gamma^1}(\mu)}%
{C_{\gamma^2 \gamma^3}(\mu)}
\nonumber\\
&{}=
\frac{C_{\gamma_5^{\text{AC}} \gamma^2 \gamma^3}(\mu)}%
{C_{\gamma_5^{\text{HV}} \gamma^2 \gamma^3}(\mu)} =
\frac{C_{\gamma^2 \gamma^3}(\mu)}%
{C_{\gamma^0 \gamma^1}(\mu)} = 1\,.
\label{Match:Ratios}
\end{align}
In particular,
$C_{\gamma_\bot \rlap{\scriptsize/}v}(\mu) =
C_{\gamma_\bot^{[\alpha} \gamma_\bot^{\beta]}}(\mu)$.
In the following we shall consider only the matching coefficients
for the first 4 Dirac structures in~(\ref{Match:Gamma}).

In order to find the coefficients $C_\Gamma(\mu)$,
we equate matrix elements of the left- and right-hand side
of~(\ref{Intro:match})
from the heavy quark with momentum $p=mv+k$
to the light quark with momentum $k_q$:
\begin{align}
&{<}q(k_q)|j(\mu)|Q(mv+k){>} = {}
\nonumber\\
&C_\Gamma(\mu) {<}q(k_q)|\tilde{\jmath}(\mu)|Q_v(k){>}
+ \mathcal{O}\left(\frac{k,k_q}{m}\right)\,.
\label{Match:match}
\end{align}
The on-shell matrix elements are
\begin{align}
{<}q(k_q)|j(\mu)|Q(p){>} ={}&
\bar{u}_q(k_q) \Gamma(p,k_q) u(p)
\nonumber\\
&{}\times Z_j^{-1}(\mu) Z_Q^{1/2} Z_q^{1/2}\,,
\nonumber\\
{<}q(k_q)|\tilde{\jmath}(\mu)|Q_v(k){>} ={}&
\bar{u}_q(k_q) \tilde{\Gamma}(k,k_q) u_v(k)
\nonumber\\
&{}\times \tilde{Z}_j^{-1}(\mu) \tilde{Z}_Q^{1/2} \tilde{Z}_q^{1/2}\,,
\label{Match:onshell}
\end{align}
where $\Gamma(p,k_q)$ and $\tilde{\Gamma}(k,k_q)$ are the bare vertex functions,
and $\tilde{Z}_q$ differs from $Z_q$ because there are no $Q$ loops in HQET.
The difference between $u(mv+k)$ and $u_v(k)$ is of order $k/m$,
and can be neglected.
It is most convenient to use $k=k_q=0$,
then the $\mathcal{O}(1/m)$ term is absent.
The QCD vertex has two Dirac structures:
\begin{equation*}
\Gamma(mv,0) = \Gamma \cdot (A + B \rlap/v)\,.
\end{equation*}
This leads to
\begin{align*}
&\bar{u}(0) \Gamma(mv,0) u(mv) = \bar{\Gamma}(mv,0)\,\bar{u}(0) \Gamma u(mv)\,,\\
&\bar{\Gamma}(mv,0) = A + B\,.
\end{align*}
The HQET vertex has just one Dirac structure.
Therefore,
\begin{equation}
C_\Gamma(\mu) =
\frac{\bar{\Gamma}(mv,0) Z_j^{-1}(\mu) Z_Q^{1/2} Z_q^{1/2}}%
{\tilde{\Gamma}(0,0) \tilde{Z}_j^{-1}(\mu) \tilde{Z}_Q^{1/2} \tilde{Z}_q^{1/2}}\,.
\label{Match:main}
\end{equation}

If all flavours except $Q$ are massless,
all loop corrections to $\tilde{\Gamma}(0,0)$,
$\tilde{Z}_Q$, and $\tilde{Z}_q$ contain no scale
and hence vanish: $\tilde{\Gamma}(0,0)=1$,
$\tilde{Z}_Q=1$, $\tilde{Z}_q=1$.
The quantities $\Gamma(mv,0)$, $Z_Q$, and $Z_q$
contain a single scale $m$;
$Z_Q$ has been calculated up to 3 loops in~\cite{MR:00},
$Z_q$ in~\cite{CKS:98},
and $\Gamma(mv,0)$ in the present work~\cite{BGMPSS:10}.
The $\overline{\text{MS}}$ renormalization constants
$\tilde{Z}_j$~\cite{CG:03}
and $Z_j$~\cite{G:00} (for all $\Gamma$)
are also known to 3 loops.

If there is another massive flavour ($c$ in $b$-quark HQET),
then $\tilde{\Gamma}(0,0)$, $\tilde{Z}_Q$, and $\tilde{Z}_q$
contain a single scale $m_c$.
The first two quantities have been calculated
up to 3 loops in~\cite{GSS:06};
the last one is known from~\cite{CKS:98}.
The quantities $\Gamma(mv,0)$, $Z_Q$, and $Z_q$
now contain 2 scales, and are non-trivial functions
of $x=m_c/m$.
The renormalization constant $Z_Q$ has been calculated
in this case, up to 3 loops, in~\cite{BGSS:07}
(the master integrals appearing in this case
are discussed in Ref.~\cite{BGSS:08}).
The other two quantities are found in this work~\cite{BGMPSS:10}.

The bare on-shell QCD quantities $\bar{\Gamma}(mv,0)$, $Z_Q$, and $Z_q$
are expressed via $g_0^{(n_f)}$
(and $m_{c0}^{(n_f)}$ if it is non-zero;
we re-express it via the on-shell mass $m_c$).
They don't contain $\mu$.
The $\overline{\text{MS}}$ QCD renormalization constant $Z_j$
is expressed via $\alpha_s^{(n_f)}(\mu)$.
The bare on-shell HQET quantities
$\tilde{\Gamma}(0,0)$, $\tilde{Z}_Q$, and $\tilde{Z}_q$
are expressed via $g_0^{(n_f-1)}$ and $m_{c0}^{(n_f-1)}$
(they are trivial at $m_c=0$);
we re-express $m_{c0}^{(n_f-1)}$ via the on-shell mass $m_c$
(which is the same in both theories).
These bare quantities also don't contain $\mu$.
Finally, the $\overline{\text{MS}}$ HQET
renormalization constant $\tilde{Z}_j$
is expressed via $\alpha_s^{(n_f-1)}(\mu)$.
We re-express all the quantities in~(\ref{Match:main})
via $\alpha_s^{(n_f-1)}(\mu)$, see~\cite{CKS:98}.

From equation of motion we have
\begin{equation}
i \partial_\alpha j^\alpha = i \partial_\alpha j_0^\alpha
= m_0 j_0 = m(\mu) j(\mu)\,,
\label{Match:div}
\end{equation}
where $m(\mu)$ is the $\overline{\text{MS}}$ mass
of the heavy quark $Q$.
Taking the on-shell matrix element
between the heavy quark with $p=mv$ and the light quark with $k_q=0$
and re-expressing both QCD matrix elements via the matrix element
of the HQET current with $\Gamma=1$, we obtain~\cite{BG:95}
\begin{equation}
m C_{\rlap{\scriptsize/}v}(\mu) = m(\mu) C_1(\mu)\,.
\label{Match:mm}
\end{equation}
The ratio $m(\mu)/m$ has been calculated at three loops numerically~\cite{CS:99}
and then analytically~\cite{MR:00a}
(the analytical results~\cite{MR:00a,MR:00} were later independently confirmed
in~\cite{MMPS:07}, and then in several other papers);
for $m_c\neq0$, $m(\mu)/m$ has been found in~\cite{BGSS:07}.

The matching coefficients have been calculated up to 2 loops in~\cite{BG:95},
and to 3 loops in the present work~\cite{BGMPSS:10}.
Analytical expressions are long;
numerically, at $m_c=0$ and $\mu=m$ we have
\begin{align*}
&C_1^{(2)} = 7.55 + 1.09 = 8.64\,,\\
&C_{\rlap{\scriptsize/}v}^{(2)} = - 5.47 + 3.06 = - 2.41\,,\\
&C_{\gamma_\bot}^{(2)} = - 9.87 + 1.53 = - 8.34\,,\\
&C_{\gamma_\bot\rlap{\scriptsize/}v}^{(2)} = - 14.13 + 2.42 = - 11.70\,,\\
&C_1^{(3)} = 64.74 + 75.34 - 38.16 = 101.92\,,\\
&C_{\rlap{\scriptsize/}v}^{(3)} = - 37.25 - 10.72 + 29.74 = - 18.23\,,\\
&C_{\gamma_\bot}^{(3)} = - 88.92 - 46.34 + 45.34 = - 89.92\,,\\
&C_{\gamma_\bot\rlap{\scriptsize/}v}^{(3)} = - 123.61 - 63.57 + 63.22 = - 123.96
\end{align*}
(in the middle part of each formula,
terms with descending powers of $\beta_0^{(n_f-1)}$ are shown separately).
Naive nonabelianization~\cite{BG:95} works reasonably well.

\begin{table*}
\caption{Master integrals with 5 lines}
\label{Tab5}
\begin{tabular}{c|cccc}
&\includegraphics[scale=0.5]{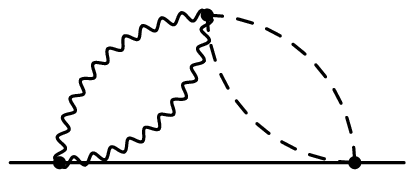}
&\includegraphics[scale=0.5]{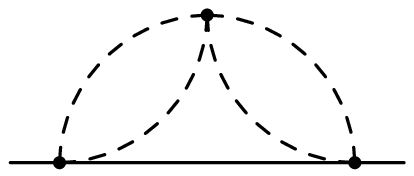}
&\includegraphics[scale=0.5]{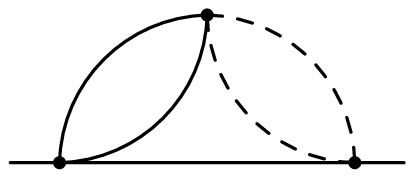}
&\includegraphics[scale=0.5]{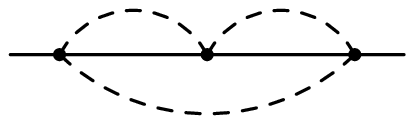}\\
&5.1, 5.1a&5.2, 5.2a&5.3, 5.3a&5.4, 5.4a\\
\hline
$\varepsilon^{-3}$&DE&DE&DE&DE\\
$\varepsilon^{-2}$&DE&DE&DE&DE\\
$\varepsilon^{-1}$&DE&DE&DE&DE\\
$1$&DE&NEW&MB&DE\\
$\varepsilon$&DE&$x$&$x$&DE\\
$\varepsilon^2$&&&&DE
\end{tabular}
\end{table*}

At $m_c\neq0$, results are expressed via the master integrals
depending on $x=m_c/m$~\cite{BGSS:08}.
Their status is summarized in the Tables~1--4 in this paper.
In the present work~\cite{BGMPSS:10},
we were able to obtain exact analytical expressions
(via harmonic polylogarithms of $x$)
for $\mathcal{O}(1)$ terms in the master integrals 5.2, 5.2a,
from the requirement of finiteness of the matching coefficients.
Therefore, the Table~3 in~\cite{BGSS:08} should be now replaced
with the following Table~\ref{Tab5}
(DE means the method of differential equations,
and MB the Mellin--Barnes representation).
Unfortunately, $\mathcal{O}(\varepsilon)$ terms
in 4 master integrals are still known only as truncated series in $x$
(the entries $x$ in the table).
Therefore, the $m_c$ corrections to the 3-loop matching coefficients
are also known only as truncated series in $x$
(or numerical approximations).

We let's apply our results to the matrix elements
between a $B$ or $B^*$ meson with momentum $p$ and the vacuum:
\begin{align}
&{<}0| \left(\bar{q} \gamma_5^{\text{AC}} Q\right)_\mu |B{>} =
- i m_B f_B^P(\mu)\,,
\label{Meson:matel}\\
&{<}0| \bar{q} \gamma^\alpha \gamma_5^{\text{AC}} Q |B{>} =
i f_B p^\alpha\,,
\nonumber\\
&{<}0| \bar{q} \gamma^\alpha Q |B^*{>} =
i m_{B^*} f_{B^*} e^\alpha\,,
\nonumber\\
&{<}0| \left(\bar{q} \sigma^{\alpha\beta} Q\right)_\mu |B^*{>} =
f_{B^*}^{T}(\mu) (p^\alpha e^\beta - p^\beta e^\alpha)\,.
\nonumber
\end{align}
The corresponding HQET matrix elements in the $v$ rest frame are
\begin{align}
&{<}0| \left(\bar{q}\gamma_5^{\text{AC}} Q_v\right)_\mu
|B(\vec{k}\,){>}\strut_{\text{nr}} = - i F(\mu)\,,
\nonumber\\
&{<}0| \left(\bar{q} \vec{\gamma} Q_v\right)_\mu
|B^*(\vec{k}\,){>}\strut_{\text{nr}} = i F(\mu) \vec{e}\,,
\label{Meson:HQET}
\end{align}
where the single-meson states are normalized by the non-relativistic condition
\begin{equation*}
\strut_{\text{nr}}{<}B(\vec{k}\,')|B(\vec{k}\,){>}\strut_{\text{nr}}
= (2\pi)^3 \delta(\vec{k}\,'-\vec{k}\,)\,.
\end{equation*}
These two matrix elements are characterized
by a single hadronic parameter $F(\mu)$
due to the heavy-quark spin symmetry.
From~(\ref{Match:div}) we have~\cite{BG:95}
\begin{equation}
\frac{f_B^P(\mu)}{f_B} = \frac{m_B}{m(\mu)}\,,
\label{Meson:mm}
\end{equation}
where we may replace $m_B$ by the on-shell $b$-quark mass $m$,
neglecting power corrections.

Our main result is the ratio $f_{B^*}/f_B$.
At $m_c=0$
\begin{align}
&\frac{f_{B^*}}{f_B} = 1
- \frac{1}{2} C_F \frac{\alpha_s^{(4)}(m)}{\pi} + {}
\nonumber\\
&\left(C_F r_F + C_A r_A + T_F n_l r_l + T_F r_h\right)
C_F \left(\frac{\alpha_s^{(4)}(m)}{\pi}\right)^2
\nonumber\\
&{} + \bigl(C_F^2 r_{FF} + C_F C_A r_{FA} + C_A^2 r_{AA}
+ C_F T_F n_l r_{Fl}
\nonumber\\
&{} + C_F T_F r_{Fh} + C_A T_F n_l r_{Al} + C_A T_F r_{Ah}
\nonumber\\
&{} + T_F^2 n_l^2 r_{ll} + T_F^2 n_l r_{lh} + T_F^2 r_{hh}\bigr)
C_F \left(\frac{\alpha_s^{(4)}(m)}{\pi}\right)^3
\nonumber\\
&{} + \mathcal{O}\left(\alpha_s^4,\frac{\Lambda}{m}\right)\,,
\label{Currents:ff}
\end{align}
where
\begin{align*}
&r_F = \frac{1}{3} \pi^2 \log 2 - \frac{1}{2} \zeta_3
    - \frac{4}{9} \pi^2 + \frac{31}{48}\,,\\
&r_A = -  \frac{1}{6} \pi^2 \log 2 + \frac{1}{4} \zeta_3
    + \frac{1}{6} \pi^2 - \frac{263}{144}\,,\\
&r_l = \frac{19}{36}\,,\quad
r_h = \frac{1}{9} \pi^2 - \frac{41}{36}\,,\displaybreak\\
&r_{FF} = - \frac{8}{3} a_4 - \frac{1}{9} \log^4 2
    - \frac{2}{9} \pi^2 \log^2 2\\
&{} + \frac{19}{6} \pi^2 \log 2 + \frac{25}{12} \zeta_5
    - \frac{1}{9} \pi^2 \zeta_3 + \frac{11}{8} \zeta_3\\
&{} - \frac{43}{1080} \pi^4 - \frac{43}{24} \pi^2 - \frac{289}{192}\,,\\
&r_{FA} = - \frac{20}{9} a_4 - \frac{5}{24} \log^4 2
    - \frac{5}{27} \pi^2 \log^2 2\\
&{} + \frac{305}{108} \pi^2 \log 2 - \frac{115}{48} \zeta_5
    + \frac{1}{12} \pi^2 \zeta_3 - \frac{899}{144} \zeta_3\\
&{} + \frac{817}{12960} \pi^4 - \frac{2233}{648} \pi^2 + \frac{4681}{864}\,,\\
&r_{AA} = \frac{16}{9} a_4 + \frac{2}{27} \log^4 2
    + \frac{4}{27} \pi^2 \log^2 2\\
&{} - \frac{119}{54} \pi^2 \log 2 + \frac{5}{6} \zeta_5
    - \frac{11}{144} \pi^2 \zeta_3 + \frac{343}{144} \zeta_3\\
&{} - \frac{17}{3240} \pi^4 + \frac{2839}{1728} \pi^2 - \frac{48125}{5184}\,,\\
&r_{Fl} = \frac{16}{9} a_4 + \frac{2}{27} \log^4 2
    + \frac{4}{27} \pi^2 \log^2 2\\
&{} - \frac{28}{27} \pi^2 \log 2 + \frac{25}{9} \zeta_3
    - \frac{11}{324} \pi^4 + \frac{179}{162} \pi^2 - \frac{815}{864}\,,\\
&r_{Fh} = - \frac{32}{9} a_4 - \frac{4}{27} \log^4 2
    + \frac{4}{27} \pi^2 \log^2 2\\
&{} + \frac{46}{27} \pi^2 \log 2 + 5 \zeta_3
    - \frac{1}{162} \pi^4 - \frac{1439}{1080} \pi^2 - \frac{119}{36}\,,\\
&r_{Al} = \frac{8}{9} a_4 - \frac{1}{27} \log^4 2
    - \frac{2}{27} \pi^2 \log^2 2\\
&{} + \frac{14}{27} \pi^2 \log 2 - \frac{13}{18} \zeta_3
    + \frac{13}{3240} \pi^4 - \frac{17}{72} + \frac{422}{81}\,,\\
&r_{Ah} = \frac{16}{9} a_4 + \frac{2}{27} \log^4 2
    - \frac{2}{27} \pi^2 \log^2 2\\
&{} - \frac{86}{27} \pi^2 \log 2 + \frac{55}{48} \zeta_5
    - \frac{31}{144} \pi^2 \zeta_3 + \frac{43}{36} \zeta_3\\
&{} + \frac{8}{405} \pi^4 + \frac{577}{270} \pi^2 - \frac{1121}{648}\,,\\
&r_{ll} = - \frac{1}{27} \pi^2 - \frac{203}{324}\,,\quad
r_{lh} = \frac{5}{81} \pi^2 - \frac{101}{162}\,,\\
&r_{hh} = - \frac{8}{9} \zeta_3 + \frac{8}{405} \pi^2 + \frac{277}{324}
\end{align*}
($a_4=\mathop{\mathrm{Li}}\nolimits_4(1/2)$).
The result for $f^T_{B^*}(m)/f_{B^*}$ is similar.

Numerically,
\begin{align*}
&\left(\frac{f_{B^*}}{f_B}\right)^{(2)} = - 4.40 - 1.97 = - 6.37\,,\\
&\left(\frac{f_{B^*}^T(m_b)}{f_{B^*}}\right)^{(2)} = - 4.26 + 0.89 = - 3.37\,,\\
&\left(\frac{f_{B^*}}{f_B}\right)^{(3)} = - 51.67 - 42.21 + 16.33 = - 77.55\,,\\
&\left(\frac{f_{B^*}^T(m_b)}{f_{B^*}}\right)^{(3)} = - 34.69 - 22.91 + 19.07\\
&{} = - 38.53\,.
\end{align*}
Naive nonabelianization~\cite{BG:95} works reasonably well.

Asymptotics of the perturbative coefficients for the matching coefficients
at a large number of loops $l\gg1$ have been investigated in Ref.~\cite{CGM:03}
in a model-independent way.
The results contain three unknown normalization constants $N_{0,1,2}\sim1$.
The asymptotics of the perturbative coefficients for $f_{B^*}/f_B$
contain $N_0$ and $N_2$;
in the case of $m/\hat{m}$ it contains only $N_0$:
\begin{align}
&\left(\frac{f_{B^*}}{f_B}\right)^{(n+1)}_{L=-5/3} =
- \frac{14}{27} \biggl\{ 1 + \mathcal{O}\left(\frac{1}{n}\right)
\nonumber\\
&{} + \frac{2}{7} \left(\frac{50}{3} n\right)^{-9/25}
\left[1 + \mathcal{O}\left(\frac{1}{n}\right) \right]
\frac{N_2}{N_0} \biggr\}
\nonumber\\
&{}\times\left(\frac{m}{\hat{m}}\right)^{(n+1)}_{L=-5/3}\,.
\label{asympt}
\end{align}
The coefficient of $N_2/N_0$ is about $0.08$ at $n=2$,
and it seems reasonable to neglect this contribution.
Neglecting also $1/n$ corrections, we obtain~\cite{CGM:03}
\begin{equation*}
\left(\frac{f_{B^*}}{f_B}\right)^{(3)}_{L=-5/3} =
- \frac{14}{27} \cdot 56.37 = - 29.23\,.
\label{predict}
\end{equation*}
Our exact result $-37.787$
agrees with this prediction reasonably well.
However, $1/n$ corrections are large and tend to break this agreement.
It is natural to expect that $1/n^2$ (and higher) corrections
are also substantial at $n=2$.

\textbf{Acknowledgements}.
I am grateful to S.~Bekavac, P.~Marquard, J.H.~Piclum, D.~Seidel,
M.~Steinhauser for collaboration on~\cite{BGMPSS:10},
and to D.J.~Broadhurst, K.G.~Chetyrkin, D.V.~Shirkov, V.A.~Smirnov
for useful discussions.

\end{document}